\documentclass[10pt,conference]{IEEEtran}
\pdfoutput=1

\usepackage[utf8]{inputenc}
\usepackage{amsmath,amssymb,amsfonts}
\usepackage{graphicx}
\usepackage{xcolor}
\usepackage{cite}
\usepackage{url}
\usepackage{placeins}
\usepackage{tabularx}
\usepackage{algorithm}
\usepackage{algorithmic}
\usepackage{subcaption}

\begin{document}
\title{ AoI-Aware Resource Allocation with Deep Reinforcement Learning for HAPS-V2X Networks  \\
       }

\author{
\IEEEauthorblockN{Ahmet Melih İnce\IEEEauthorrefmark{1}, Ayşe Elif Canbilen\IEEEauthorrefmark{1}, Halim Yanikomeroglu\IEEEauthorrefmark{7}}
\IEEEauthorblockA{\IEEEauthorrefmark{1}Konya Technical University, Konya, 42250, Türkiye \\
Emails: \{e228221001009, aecanbilen\}@ktun.edu.tr}
\IEEEauthorblockA{\IEEEauthorrefmark{7}Carleton University, Ottawa, ON K1S 5B6, Canada \\
Email: halim@sce.carleton.ca}
}

\maketitle

\begin{abstract}
Sixth-generation (6G) networks are designed to meet the hyper-reliable and low-latency communication (HRLLC) requirements of safety-critical applications such as autonomous driving. Integrating non-terrestrial networks (NTN) into the 6G infrastructure brings redundancy to the network, ensuring continuity of communications even under extreme conditions. In particular, high-altitude platform stations (HAPS) stand out for their wide coverage and low latency advantages, supporting communication reliability and enhancing information freshness, especially in rural areas and regions with infrastructure constraints.
In this paper, we present reinforcement learning-based approaches using deep deterministic policy gradient (DDPG) to dynamically optimize the age-of-information (AoI) in HAPS-enabled vehicle-to-everything (V2X) networks. The proposed method improves information freshness and overall network reliability by enabling independent learning without centralized coordination. The findings reveal the potential of HAPS-supported solutions, combined with DDPG-based learning, for efficient AoI-aware resource allocation in platoon-based autonomous vehicle systems.
\end{abstract}

\begin{IEEEkeywords}
HAPS, V2X, AoI, Multi-Agent Reinforcement Learning, 6G.
\end{IEEEkeywords}

\section{Introduction}

Intelligent transportation and autonomous driving systems have emerged as key components of modern wireless communication research \cite{9815183}. The increasing need for \emph{hyper-reliable, low-latency communication (HRLLC)} in vehicular networks requires the development of advanced communication frameworks that ensure efficient data exchange between vehicles, infrastructure, and cloud/edge nodes. With the advent of sixth-generation (6G) networks, these systems are expected to support massive connectivity, ultra-low latency, and high data rates\cite{noor20226g}. However, several challenges remain, particularly in environments with limited or unreliable terrestrial infrastructure, such as remote areas, disaster-stricken zones, and ultradense urban environments \cite{10488039}. 
For instance, while platoon-based networks require robust intra- and inter-platoon communications to maintain synchronized movement and dynamic adaptation to road conditions, keeping data fresh in such a system is challenging due to varying network conditions, mobility patterns, and channel uncertainties.

In this regard, a critical metric in vehicular network research is \emph{Age of Information (AoI)}, which quantifies the timeliness or ``freshness'' of data received by a target node \cite{9562190}. Unlike traditional network performance metrics such as throughput and latency, AoI provides direct insight into how current the received information is, making it particularly relevant for applications such as autonomous vehicle control, collision avoidance, and real-time traffic management\cite{10443065}. Ensuring low AoI is essential for maintaining situational awareness in highly dynamic vehicular environments, where even minor delays in information updates can lead to significant safety risks.

Integrating HAPS into vehicular communication networks offers a promising solution to overcome coverage limitations and challenges related to AoI\cite{10794340}. Operating at altitudes of approximately 20\,km, HAPS provides wide-area connectivity, strong line-of-sight (LoS) links, and low-latency communication, effectively complementing terrestrial and satellite-based networks\cite{10387578, 9815183}. By acting as aerial base stations or relays, HAPS can enhance the reliability and freshness of data in vehicle-to-everything (V2X) networks, particularly in infrastructure-constrained areas \cite{kurt2021vision}.

To further improve resource allocation and minimize AoI in HAPS-assisted V2X networks, deep reinforcement learning (DRL) techniques have shown remarkable potential. DRL enables autonomous and dynamic decision-making, allowing vehicular agents to optimize real-time communication strategies \cite{9772280}. Among various DRL approaches, deep deterministic policy gradient (DDPG) and its multi-agent extension, multi-agent DDPG (MADDPG), have demonstrated effectiveness in handling complex high-dimensional control problems \cite{ye2019deep}. While DDPG provides a straightforward, independent learning mechanism, its limited adaptation to external interference affects its effectiveness. Fully decentralized MADDPG (FD-MADDPG) on the other hand, offers a more scalable and adaptable solution, allowing multiple agents to learn in parallel without centralized dependency. The decentralized framework, combined with HAPS integration, positions FD-MADDPG as a promising approach for optimizing AoI-aware communication in next-generation vehicular networks.

This study focuses on optimizing AoI-aware resource allocation in HAPS-enabled V2X networks. In particular, we 
propose a DRL-based framework using DDPG to optimize AoI, improving communication efficiency and autonomous decision-making. By integrating platoon-based vehicular coordination, we develop a resource allocation model that enhances intra- and inter-platoon data exchange, ensuring low AoI and stable connectivity. Our model leverages HAPS as an aerial relay to extend network coverage, enhance reliability, and provide seamless connectivity in infrastructure-limited scenarios. The provided simulation results validate the effectiveness of our approach, demonstrating significant improvements in AoI reduction and network reliability.

The rest of this paper is structured as follows. Section~\ref{sec:ProblemForm} presents the system model and problem formulation, detailing the role of HAPS in the V2X framework. Section~\ref{sec:DRL-Approaches} describes the DRL-based resource allocation methods, while Section~\ref{sec:Results} provides the simulation results. Finally, Section~\ref{sec:Conclusions} concludes the study and outlines future research directions.

All symbols used throughout this paper are summarized in Table I and they are also defined in first appearance for clarity and consistency.

\begin{figure*}[t]  
\centering
\includegraphics[width=10cm,height=5cm]{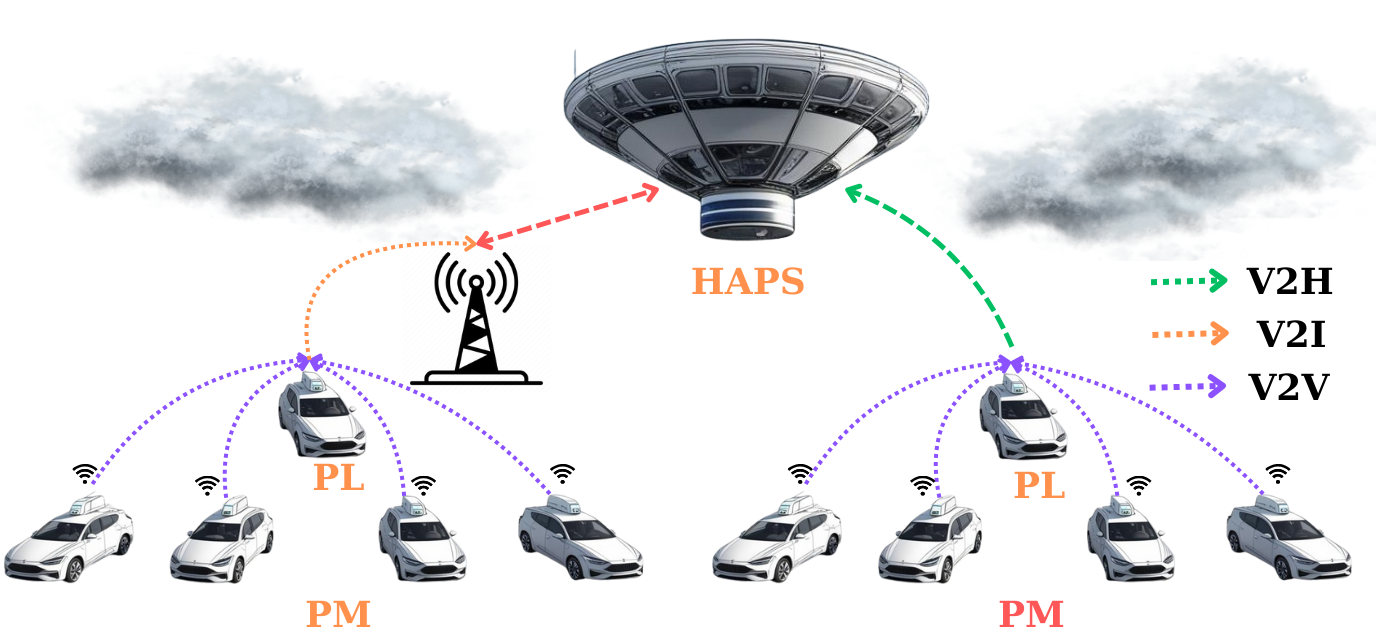}
\caption{The considered HAPS-V2X system model.} 
\label{fig:Model}
\end{figure*}

\section{System Model and Problem Formulation}\label{sec:ProblemForm}
The HAPS-enabled V2X architecture considered in this study is represented in Fig.~\ref{fig:Model} at the top of the next page. This vehicular network consists of multiple autonomous vehicle platoons, each led by a platoon leader (PL) responsible for both intra- and inter-platoon communications. The system supports three primary communication modes, namely vehicle-to-infrastructure (V2I), vehicle-to-vehicle (V2V), and vehicle-to-HAPS (V2H). In V2I communication, the PL connects to roadside units (RSU) or cellular base stations to exchange critical data \cite{10103832}. V2V communication facilitates direct short-range data sharing within a platoon, ensuring coordinated movement and operational efficiency. Lastly, V2H communication leverages HAPS, which operates at approximately 20 km altitude, to extend coverage beyond terrestrial infrastructure and enhance network reliability, particularly in remote or infrastructure-limited areas, by providing real-time updates and connectivity resilience.

\subsection{Channel Models}
The network operates over a set of $K$ orthogonal sub-channels, assuming that orthogonal frequency division
multiplexing (OFDM) is utilized. The channel gains for the $j$th PL communicating with infrastructure, other vehicles in the platoon, i.e., platoon members, and HAPS through the sub-channel $k\in\mathcal{K}=\{1,\dots,K\}$ at a time slot $t$ are denoted as $h_{j,I}^t[k]$, $h_{j,v}^t[k]$, and $h_{j,H}^t[k]$, respectively. 

Given the strong likelihood of a LoS link to HAPS, the V2H channel following a Rician fading model can be expressed as
\begin{equation}
h^t_{j,H}[k] = 10^{-\mathcal{PL}/20} \left( \sqrt{p_{L}} a^t_j + \sqrt{p_{N}} h^t_j[k] \right),
\label{eq1}
\end{equation}
where $\mathcal{PL}$ denotes path loss defined by considering the scintillation loss caused by rapid fluctuations
of the received signal, the attenuation
stemming from atmospheric gases, the clutter loss, the shadow fading, and the free-space path loss, all specified in 3GPP standards \cite[Tables 6.6.2-1–6.6.2-3]{3gpp}.
Additionally, $p_L$ and $p_N$ in \eqref{eq1} represent LoS and non-LoS probabilities, respectively\footnote{Details on the calculation of these terms can be found in \cite{9518388}.}. The term
$a^t_j$ indicates the deterministic LoS component, and $h^t_j[k]$ captures small-scale fading. 

On the other side, the gain of the channels between PL $j$ and its follower $v$, $h_{j,v}$, or infrastructure, $h_{j,I}$, in $k$th sub-channel during the $t$th coherence time is given by \cite{9562190}
\begin{equation}
    h_{j,m}^t[k] = \alpha_{j,m}^t g_{j,m}^t[k], \;\; m \in \{v,I\},
    \label{eq:channel_gain}
\end{equation}
where \( \alpha_{j,m}^t \) represents the large-scale fading effect based on path loss and shadowing, while \( g_{j,m}^t[k] \) indicates the small-scale fading. 

\begin{table}[htbp]
    \caption{List of Notations}
    \centering
    \renewcommand{\arraystretch}{1.2}
    \begin{tabular}{|c|l|}
        \hline
        \textbf{Symbol} & \textbf{Description} \\ \hline
        $A_j^t$ & AoI for agent $j$ at time $t$ \\ \hline
        $p_j^t[k]$ & Transmission power of agent $j$ at time $t$ on channel $k$ \\ \hline
        $C_{j,I}^t[k]$ & Capacity of V2I communication \\ \hline
        $C_{j,v}^t[k]$ & Capacity of V2V communication \\ \hline
        $C_{j,H}^t[k]$ & Capacity of V2H communication \\ \hline
        $\theta_j^t$ & Communication mode selection (V2I, V2V, V2H) \\ \hline
        $\beta_j^t[k]$ & Subchannel selection status \\ \hline
        $\sigma^2$ & Noise power \\ \hline
        $\Delta t$ & Time slot duration \\ \hline
        $\zeta_j$ & Minimum data transmission requirement \\ \hline
        $F(\cdot)$ & Function limiting energy consumption \\ \hline
        $G(\cdot)$ & Step function \\ \hline
        $I_j^t[k]$ & Interference on channel $k$ for agent $j$ at time $t$ \\ \hline
        $p_L$ & LoS probability \\ \hline
        $p_N$ & NLoS probability \\ \hline
        $\mathcal{PL}$ & Overall path loss \\ \hline
        $h_j^t[k]$ & Channel gain \\ \hline
        
        $P_{\text{max}}$ & Maximum transmission power allowed (in dBm) \\ \hline
        $T$ & Number of time slots\\ \hline
        $\kappa_1, \kappa_2, ...$ & Weights used in the reward function \\ \hline
        $N$ & Total number of vehicles/agents in the simulation \\ \hline
        $s_j^t$ & State observed by agent $j$ at time $t$ \\ \hline
        $a_j^t$ & Action taken by agent $j$ at time $t$ \\ \hline
        $r_j^t$ & Local reward received by agent $j$ \\ \hline
        $r^t$ & Global reward shared among agents \\ \hline
        $\pi_j$ & Policy of agent $j$\\ \hline
        $\gamma$ & Discount factor in reinforcement learning\\ \hline
        $D_j$ & Replay buffer for experience storage \\ \hline
        
    \end{tabular}
    \label{tab:notations}
\end{table}

\subsection{Problem Formulation}
In the proposed system, each PL maintains an AoI metric $A_j^t$ that quantifies data freshness. The AoI is updated for $j$th PL at time $(t+1)$ based on the selected communication mode denoted by $\theta_j^t$ as follows:
\begin{equation} 
A_{j}^{t+1} = 
\begin{cases} 
    \Delta t, & \text{if } \theta_j^t = 0 \text{ and } C_{j,I}^t[k] \geq C_{j,I}^{\min}, \\
    \Delta t, & \text{if } \theta_j^t = 2 \text{ and } C_{j,H}^t[k] \geq C_{j,H}^{\min}, \\
    A_{j}^t + \Delta t, & \text{otherwise}.
\end{cases}
\end{equation}
Here, $C_{j,I}^{\min}$ and $C_{j,H}^{\min}$ denote the minimum required capacities for successful V2I and V2H transmissions, respectively. If no valid update is received, the AoI increases, reflecting data staleness. 


Let \(\beta_{j,k}^t \in \{0,1\}\) indicate whether sub-channel \(k\) is assigned to $j$th PL at time \(t\), and  \(\theta_j^t \in \{0,1,2\}\). Here, \(\theta_j^t = 0\) corresponds to V2I, \(\theta_j^t = 1\) corresponds to V2V, and \(\theta_j^t = 2\) corresponds to V2H. The transmission power allocated by PL~\(j\) on sub-channel \(k\) is denoted as \(p_j^t[k]\). 
Then, the achievable capacity of PL~\(j\) on sub-channel \(k\) for V2H mode is written by
\begin{equation}
C_{j,H}^t[k] = \log_2 \left( 1 + \frac{\delta(\theta_j^t - 2) \beta_{j,k}^t p_j^t[k] h_{j,H}^t[k]}{I_j^H[k] + \sigma^2} \right),
\end{equation}
where \(\delta(\cdot)\) is the indicator function ensuring that the capacity formula is applied only when PL~\(j\) operates in V2H mode (\(\theta_j^t = 2\)). The term \(\sigma^2\) refers to the thermal noise power, and  \(I_j^H[k]\) denotes the total interference power from other PL using the same sub-channel and can be given as follows
\begin{equation}
I_j^H[k] = \sum_{j' \neq j} \beta_{j',k}^t p_{j'}^t[k] h_{j',H}^t[k].
\end{equation}

A similar capacity formulation applies for V2I and V2V links, replacing the corresponding channel gain \(h_{j,H}^t[k]\) with \(h_{j,I}^t[k]\) for V2I and \(h_{j,v}^t[k]\) for V2V links, as defined in \eqref{eq:channel_gain}, while adjusting the interference term and \(\delta(\cdot)\) function accordingly.

We aim to minimize the average AoI and power consumption for every platoon while respecting capacity constraints and ensuring reliable data delivery among the platoon members. Accordingly, the multi-objective optimization problem formulated for platoon $j$ can be defined as:
\begin{align*}
\min_{\beta, \theta, p } \Bigg\{ & \frac{1}{T} \sum_{t=1}^{T} A_j^t - \Pr \left\{ \sum_{t=1}^{T} \sum_{k \in \mathcal{K}} \underset{v}{\min} \left\{ C_{j,v}^t[k] \right\} \Delta t \geq \zeta_j \right\} \\
& +\frac{1}{T} \sum_{t=1}^{T} \sum_{k \in \mathcal{K}} p_j^t[k] \Bigg\},
\end{align*}
   \begin{align} \textbf{s.t.} \;\;
   C1: & \, C^t_{j,H}[k] \geq C_{j,H}^{\text{min}}, \quad \forall j \in \mathcal{P}, \, \forall k \in \mathcal{K}, \,  \nonumber\\
    C2: & \, C^t_{j,I}[k] \geq C_{j,I}^{\text{min}}, \quad \forall j \in \mathcal{P}, \, \forall k \in \mathcal{K}, \,  \nonumber\\
       C3: & \, \beta^t_{j,i}\in \{0, 1\}, \; \theta_j \in \{0, 1, 2\}, \; \forall j \in \mathcal{P}, \, i \in \{I, V, H\},  
       \nonumber\\
   C4: & \sum_{k \in \mathcal{K}} \beta^t_{j,i} \leq 1, \quad \forall j \in \mathcal{P}, \, \forall t \in \mathcal{\mathbb{N}},
   \nonumber\\
   C5: & \, p^t_j[k] \leq p^{\text{max}}_j, \quad \forall j \in \mathcal{P}, \, \forall k \in \mathcal{K}, \label{eq17}
   \end{align}
where  $\zeta_j $ is the size of the cooperative awareness
message (CAM), while $\mathcal{P}=\{1,2,...,P\}$. The objective function given in \eqref{eq17} consists of minimizing the average AoI, represented by \( \frac{1}{T} \sum_{t=1}^{T} A_j^t \), and ensuring the probability that the total communication capacity overall time slots and channels meets or exceeds the minimum data requirement, i.e., \( \Pr \left\{ \sum_{t=1}^{T} \sum_{k \in \mathcal{K}} \underset{v}{\min} \left\{ C_{j,v}^t[k] \right\} \Delta t \geq \zeta_j \right\} \). Furthermore, optimization aims to minimize total power consumption in all channels and time slots, given by \( \frac{1}{T} \sum_{t=1}^{T} \sum_{k \in \mathcal{K}} p_j^t[k] \).

In \eqref{eq17}, the first constraint ensures that the communication capacity \( C^t_{j,i}[k] \) satisfies a minimum value \( C_{j,i}^{\text{min}} \). The second and third constraints force each agent to choose a valid communication mode and use only one sub-channel at any given time. With the last constraint, the transmission power for each agent is limited to a maximum value \( p^{\text{max}}_j \).

\section{DRL Approaches for AoI Optimization}\label{sec:DRL-Approaches}

This section explores two reinforcement learning approaches to optimize AoI in HAPS-assisted V2X networks. The first, DDPG, follows a single-agent paradigm, where each PL independently optimizes its AoI based on local observations. While this allows decentralized decision-making, it does not inherently account for inter-agent dependencies, leading to suboptimal resource allocation in congested networks. Due to the lack of inter-agent coordination, DDPG struggles to adapt to dynamic interference patterns, resulting in performance degradation and slower convergence.

The second approach, FD-MADDPG, extends DDPG to a multi-agent reinforcement learning framework, allowing multiple PL to learn concurrently without explicit coordination. Unlike traditional MADDPG, which employs a centralized critic, FD-MADDPG eliminates the need for centralized training. This enables real-time decisions based solely on local observations, improving scalability and robustness in large-scale V2X networks. Using independent learning strategies, FD-MADDPG is known to achieve faster convergence and lower AoI, particularly in high-mobility and dense-vehicle scenarios. In addition, it enables dynamically adjusting transmission and resource allocation strategies based on environmental variations. 
This makes FD-MADDPG more efficient in handling network congestion and spectral efficiency. 

In both approaches, each PL acts as an agent that interacts with the vehicle environment by observing situations and taking the necessary actions according to its predefined policy. Therefore, at any time $t$, each PL $j$ observes the state space, which is given by 
\begin{align}
    s^t_j = [h^t_{j,v}[k], h^t_{j,I}[k], h^t_{j,H}[k], I^{t-1}_j[k], A^t_j, \zeta^r_j, T^r_j]. \label{eq7}
\end{align}
According to \eqref{eq7}, PLs observe not only 
the instant channel states but also the AoI and the amount of interference caused by other platoons in the previous step. In addition, $\zeta^r_j$ and $T^r_j$ in \eqref{eq7} represent the remaining intra-platoon message load and the remaining time budget, respectively. It should be noted here that each PL maintains an independent experience replay buffer $D_j = \{s_j^t, a_j^t, r_j^t, s_j^{t+1}\}$, ensuring that training remains decentralized.


\begin{algorithm}[htbp]
\caption{Training of DDPG and FD-MADDPG}
\begin{algorithmic}[1]
    \STATE Initialize actor network $\pi_{\theta_j}$ and critic network $Q_{\phi_j}$ with random weights.
    \STATE Initialize target networks $\pi_{\theta_j'}$ and $Q_{\phi_j'}$.
    \STATE Initialize experience replay buffer $D_j$.
    \FOR{each episode}
        \STATE Reset environment and receive initial state $s_j^0$.
        \FOR{each time step $t$}
            \STATE Select action $a_j^t = \pi_{\theta_j}(s_j^t) + \mathcal{N}$ (with exploration noise $\mathcal{N}$).
            \STATE Execute action $a_j^t$, observe reward $r_j^t$ and next state $s_j^{t+1}$.
            \STATE Store transition $(s_j^t, a_j^t, r_j^t, s_j^{t+1})$ in $D_j$.

            \STATE Sample minibatch from $D_j$.

            \STATE Compute target value:
            \begin{equation}
            y_j^t = r_j^t + \gamma Q_{\phi_j'}(s_j^{t+1}, \pi_{\theta_j'}(s_j^{t+1})). \nonumber
            \end{equation}

            \STATE Update critic network by minimizing loss.
            \begin{equation}
            L(\phi_j) = \frac{1}{N} \sum \left(Q_{\phi_j}(s_j^t, a_j^t) - y_j^t\right)^2. \nonumber
            \end{equation}

            \STATE Update actor network:
            \begin{equation}
            \triangledown_{\theta_j} J_j = \mathbb{E} \left[ \triangledown_{\theta_j} \pi_j (a_j | s_j) \triangledown_{a_j} Q_{\phi_j} (s_j, a_j) \right]. \nonumber
            \end{equation}

            \STATE Soft update target networks.
        \ENDFOR
    \ENDFOR


\end{algorithmic}
\end{algorithm}

Next, the local reward function for both DDPG and FD-MADDPG is defined as follows
\begin{align}
r_j^t = 
& - \kappa_1 F \left( p_j^{t} \right) - \kappa_2 A_j^{t} - \kappa_3 G \left( C_{j,I}^t - C_{j,I}^{\text{min}} \right) \notag \\
& - \kappa_4 G \left( C_{j,H}^t - C_{j,H}^{\text{min}} \right),
\end{align}
where $F(\cdot)$ penalizes high transmission power while restricting it to the same range as the other components, and the stepwise function $G(\cdot)$ ensures that minimum capacity constraints are met \cite{9562190}. However, a key distinction between DDPG and FD-MADDPG lies in how they optimize their reward function. In particular, DDPG utilizes a centralized critic that considers the overall interference in the environment. 
In contrast, FD-MADDPG operates in a fully decentralized manner, where each agent optimizes its reward independently. 
Considering that, Algorithm 1 provides a structured approach to training both DDPG and FD-MADDPG. While DDPG benefits from a global interference-aware reward function as outlined in Algorithm-1, FD-MADDPG relies solely on local observations and independent reward updates. 

Both DDPG and FD-MADDPG employ a single-critic approach per agent that can be described by:
\begin{enumerate}
     \item Each agent maintains an independent \emph{critic network} $Q_{\phi_j}$, updated using its own local rewards $r_j^t$.
    \item Each agent maintains an \emph{actor network} $\pi_{\theta_j}$, which determines optimal actions based on local state observations.
\end{enumerate}

Furthermore, to maximize the reward for each agent, the policy gradient is updated for both methods as
\begin{align} 
\triangledown_{\theta_j} J_j &= \mathbb{E}_{s_j, a_j \sim D_j} \left[ \triangledown_{\theta_j} \pi_j (a_j | s_j) \triangledown_{a_j} Q_{\phi_j} (s_j, a_j) \right], \label{eq12}
\end{align}
in which $\pi_j (a_j | s_j)$
denotes the policy function for $j$th agent, which defines the probability of selecting action \(a_j\) given the state \(s_j\). The term \( Q_{\phi_j}(s_j, a_j) \) refers to the action-value function approximated by the critic network parameterized by \( \phi_j \), which estimates the expected cumulative discounted reward obtained by executing action \( a_j \) in state \( s_j \), and subsequently following the agent’s policy \( \pi_j \). This function serves as a baseline for the policy gradient and facilitates the evaluation of action quality within continuous action domains. Besides, $\mathbb{E}_{s, a \sim D}$ in \eqref{eq12} indicates the expected value over state-action pairs sampled from the experience replay buffer. 

Finally, it should be noted that the critic network is updated by minimizing the following loss function:
\begin{align}
L(\phi_j) = \frac{1}{N} \sum_{i=1}^{N} \left(Q_{\phi_j}(s_j^t, a_j^t) - y_j^t\right)^2,
\end{align}
where $y_j^t = r_j^t + \gamma Q_{\phi_j'}(s_j^{t+1}, \pi_{\theta_j'}(s_j^{t+1}))$ with reward discount factor $\gamma$.

\section{Numerical Results}\label{sec:Results}

This section presents simulation results for the proposed HAPS-V2X network consisting of a HAPS, an RSU, and five PLs with six followers each in an urban area.
Specifically, two DRL-based approaches are utilized in this scenario to optimize resource allocation with minimum AoI and are compared in terms of average AoI value and reward function convergence. The values of the other parameters used for the simulations are determined as shown in Table II.

\begin{figure}[!t]  \hspace{-.5cm}
\includegraphics[width=9.8cm,height=7cm]{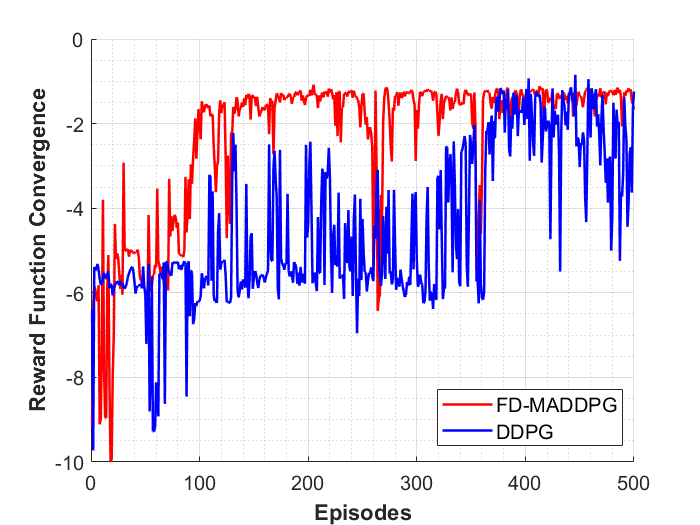}\\
\caption{Comparison of the reward function convergence.}
\label{fig:compareAlgorithm}
\end{figure}

 Fig.~\ref{fig:compareAlgorithm} presents a comparison of the reward function convergence for the utilized methods. According to this figure,
 FD-MADDPG clearly converges faster to a higher and more precise reward value compared to the DDPG approach. 
 This is due to the fact that each PL in the FD-MADDPG learns independently using only its local observations, which accelerates convergence in multi-agent environments.
 Accordingly, considering also the long-term fluctuations, it can be concluded from this figure that the learning process of DDPG is longer and more challenging than that of FD-MADDPG for the considered HAPS-V2X scenario.

\begin{table}[htbp]
\centering
\caption{Initial Parameters for the Simulation}
\label{tab:initial_parameters}
\resizebox{0.47\textwidth}{!}{%
\begin{tabular}{|l|p{2cm}|}
\hline
\textbf{Parameter Description}                                   & \textbf{Value}         \\ \hline
The distance between platoon vehicles                 & 25 m             \\ \hline
Max. transmission power of PL                   & 30 dBm                \\ \hline
Min. required data rate for V2I                 & 540 kbps              \\ \hline
Available total bandwidth                                        & 180 kHz               \\ \hline
Data size in inter-vehicle communication                   & 4000 Bytes            \\ \hline
Batch size used for training                                     & 64                    \\ \hline
Reward discount factor (\( \gamma \))                           & 0.99                  \\ \hline
Standard deviation of noise (\( \sigma \))                & 0.3 dB                \\ \hline
Actor layer dimensions              & {[}1024, 512{]}       \\ \hline
Critic layer dimensions & {[}1024, 512, 256{]}  \\ \hline
Learning rate of actor (\( \alpha \)) & 0.0001 \\ \hline
Learning rate of critic (\( \beta \)) & 0.001 \\ \hline
\end{tabular}%
}
\end{table}

\begin{figure}[!t]  \centering
\includegraphics[width=9cm,height=6.5cm]{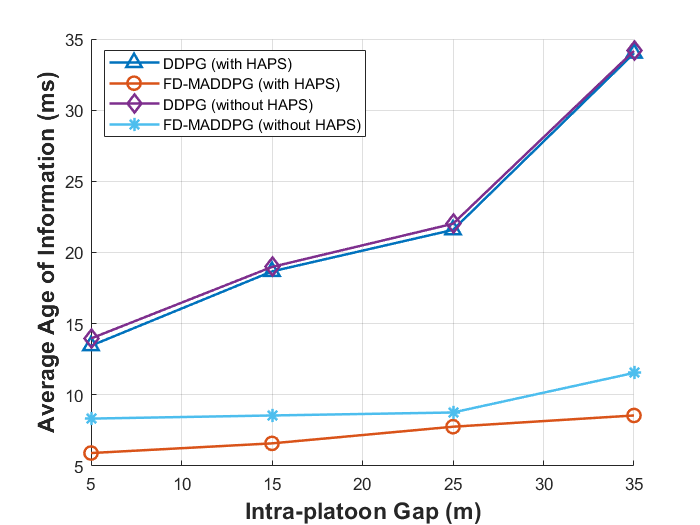}\\
\caption{Comparison of the average AoI.} 
\label{fig:AoIforAlgorithm}
\end{figure}

In Fig.~\ref{fig:AoIforAlgorithm}, FD-MADDPG is shown to be providing much lower AoI values in V2X networks compared to DDPG under similar training conditions, which can be further reduced with HAPS support. In addition, it is observed that DDPG is heavily affected by the increase in inter-platoon spacing. For example, in the HAPS-V2X scenario, the average AoI for DDPG is approximately 13 ms while for FD-MADDPG it is only 6 ms when the gap between platoons is 5 m. It is noteworthy that when the spacing between platoons is increased to 35 m, the AoI increases by only about 3 ms for FD-MADDPG, whereas it increases by almost 20 ms for DDPG. On the one hand, this shows that the decentralized solution handles interference and channel variations more effectively, keeping information fresher across the network and improving overall network reliability in HAPS-supported V2X scenarios.
On the other hand, it indicates that the information update rate of the DDPG algorithm remains relatively slow even with HAPS support.

\section{Conclusion}\label{sec:Conclusions}

This paper investigates the contribution of HAPS integration and the effectiveness of two different DRL approaches in resource allocation prioritizing information freshness in V2X networks.
Numerical evaluations have proved significant improvements in network reliability, showing that lower AoI, faster convergence, and thus, better spectrum utilization can be achieved with FD-MADDPG compared to the conventional DDPG model.
These findings highlight that HAPS can play a critical role in providing uninterrupted connectivity in time-critical scenarios, especially in environments with limited infrastructure. It should be noted here that the widespread use of solar-powered HAPS will also add a sustainable and environmentally friendly dimension to 6G networks, supporting greener communications. 

Future research can be focused on exploring energy-efficient learning strategies, adaptive reward mechanisms, and real-world deployments in large-scale vehicular networks. Additionally, investigating HAPS mobility and hybrid artificial intelligence-driven optimization can be considered to further enhance the adaptability of these networks. Comparing the performance of the proposed scheme with other recently used optimization methods, such as attention-based DRL, federated learning, or game-theoretic approaches, can be another stimulating direction.


\FloatBarrier
\bibliographystyle{ieeetr}
\bibliography{Ref}

\begin{thebibliography}{10}

\bibitem{9815183}
W.~Jaafar and H.~Yanikomeroglu, ``{HAPS-ITS}: Enabling future {ITS} services in trans-continental highways,'' {\em IEEE Communications Magazine}, vol.~60, no.~10, pp.~80--86, 2022.

\bibitem{noor20226g}
M.~Noor-A-Rahim, Z.~Liu, H.~Lee, M.~O. Khyam, J.~He, D.~Pesch, K.~Moessner, W.~Saad, and H.~V. Poor, ``{6G} for vehicle-to-everything {(V2X)} communications: Enabling technologies, challenges, and opportunities,'' {\em Proceedings of the IEEE}, vol.~110, no.~6, pp.~712--734, 2022.

\bibitem{10488039}
J.~Clancy, D.~Mullins, B.~Deegan, J.~Horgan, E.~Ward, C.~Eising, P.~Denny, E.~Jones, and M.~Glavin, ``Wireless access for {V2X} communications: Research, challenges and opportunities,'' {\em IEEE Communications Surveys \& Tutorials}, vol.~26, no.~3, pp.~2082--2119, 2024.

\bibitem{9562190}
M.~Parvini, M.~R. Javan, N.~Mokari, B.~A. Arand, and E.~A. Jorswieck, ``{AoI} aware radio resource management of autonomous platoons via multi agent reinforcement learning,'' in {\em 2021 17th International Symposium on Wireless Communication Systems (ISWCS)}, pp.~1--6, 2021.

\bibitem{10443065}
Annu and P.~Rajalakshmi, ``Towards {6G} {V2X} sidelink: Survey of resource allocation—mathematical formulations, challenges, and proposed solutions,'' {\em IEEE Open Journal of Vehicular Technology}, vol.~5, pp.~344--383, 2024.

\bibitem{10794340}
A.~M. Ince, A.~E. Canbilen, and H.~Yanikomeroglu, ``{HAPS}-enabled {V2X} architecture for hyper reliable and low-latency communication ({HRLLC}) in {6G} networks,'' in {\em International Conference on Communications, Signal Processing, and their Applications (ICCSPA)}, pp.~1--6, 2024.

\bibitem{10387578}
O.~Abbasi and H.~Yanikomeroglu, ``{UxNB}-enabled cell-free massive {MIMO} with {HAPS}-assisted sub-{THz} backhauling,'' {\em IEEE Transactions on Vehicular Technology}, vol.~73, no.~5, pp.~6937--6953, 2024.

\bibitem{kurt2021vision}
G.~K. Kurt, M.~G. Khoshkholgh, S.~Alfattani, A.~Ibrahim, T.~S. Darwish, M.~S. Alam, H.~Yanikomeroglu, and A.~Yongacoglu, ``A vision and framework for the high altitude platform station {(HAPS)} networks of the future,'' {\em IEEE Communications Surveys \& Tutorials}, vol.~23, no.~2, pp.~729--779, 2021.

\bibitem{9772280}
Q.~Ren, O.~Abbasi, G.~K. Kurt, H.~Yanikomeroglu, and J.~Chen, ``Caching and computation offloading in high altitude platform station {(HAPS)} assisted intelligent transportation systems,'' {\em IEEE Transactions on Wireless Communications}, vol.~21, no.~11, pp.~9010--9024, 2022.

\bibitem{ye2019deep}
H.~Ye, G.~Y. Li, and B.-H.~F. Juang, ``Deep reinforcement learning based resource allocation for {V2V} communications,'' {\em IEEE Transactions on Vehicular Technology}, vol.~68, no.~4, pp.~3163--3173, 2019.

\bibitem{10103832}
Q.~Ren, O.~Abbasi, G.~K. Kurt, H.~Yanikomeroglu, and J.~Chen, ``Handoff-aware distributed computing in high altitude platform station {(HAPS)}–assisted vehicular networks,'' {\em IEEE Transactions on Wireless Communications}, vol.~22, no.~12, pp.~8814--8827, 2023.

\bibitem{3gpp}
3GPP, ``Technical specification group radio access network; study on new radio {(NR)} to support non-terrestrial networks (release 15),'' {\em 3GPP TR 38.811 V15.1.0}, Jun. 2019.

\bibitem{9518388}
S.~Alfattani, W.~Jaafar, Y.~Hmamouche, H.~Yanikomeroglu, and A.~Yongaçoglu, ``Link budget analysis for reconfigurable smart surfaces in aerial platforms,'' {\em IEEE Open Journal of the Communications Society}, vol.~2, pp.~1980--1995, 2021.

\end{thebibliography}
\FloatBarrier

\end{document}